\documentclass[%
reprint,
amsmath,amssymb,
aps,
prd,
floatfix,
]{revtex4-2}

\usepackage{graphicx}% Include figure files
\usepackage{dcolumn}% eqnarray table columns on decimal point
\usepackage{bm}% bold math
\usepackage{diagbox}
\begin{document}
	
	%\preprint{APS/123-QED}
	
	\title{Higher-order NLO initial state QED radiative corrections to $e^+e^-$ annihilation revisited}% Force line breaks with \\
	
	\author{A.~B.~Arbuzov}
	\email{arbuzov@theor.jinr.ru}
	\author{U.~E.~Voznaya}%
	\email{voznaya@theor.jinr.ru}
	\affiliation{Joint Institute for Nuclear Research,             
		Joliot-Curie str. 6, Dubna, 141980, Russia \\
		Dubna state university, Universitetskaya str. 19, Dubna, 141980, 
		Russia}%Lines break automatically or can be forced with \\

	\date{\today}% It is always \today, today,
	
	\begin{abstract}
		Radiative corrections due to initial state radiation in electron-positron
		annihilation are calculated within the QED structure function approach.
		Results are shown in the next-to-leading logarithmic approximation up to
		$\mathcal{O}(\alpha^4 L^3)$ order, where $L=\ln(s/m_e^2)$ is the large logarithm.
		Several mistakes in previous calculations are corrected. The results are relevant 
		for future high-precision experiments at $e^+e^-$ colliders.
	\end{abstract}
	
	\maketitle
	
	%\tableofcontents
	
	\section{\label{sec:level1}Introduction}
	
	The physical program of future electron-positron colliders such as 
	FCCee~\cite{FCC:2018byv},  
	CEPC~\cite{CEPCStudyGroup:2018ghi},
	and ILC~\cite{ILC:2013jhg},
	foresee extremely high experimental precision in measurements of 
	$e^+e^-$ annihilation and scattering processes. In particular, it is 
	planned to collect up to $10^{12}$ events with production of $Z$ bosons
	in the so-called TeraZ operation mode~\cite{FCC:2018byv} at the $Z$ peak.
	The foreseen precision of the experimental measurements challenges for
	increasing accuracy of theoretical predictions~\cite{Jadach:2019bye}. 
	
	Computation of the complete $\mathcal{O}(\alpha^2)$ electroweak and
	even QED radiative corrections to realistic observables is still 
	a difficult problem. The QED structure function\footnote{The
		functions can be better called as QED parton distribution ones
		(QED PDFs).} approach~\cite{Kuraev:1985hb} allows taking 
	systematically into account the terms enhanced by the so-called
	large logarithm
	\begin{equation}
		L = \ln \frac{\mu_F^2}{\mu_R^2},
	\end{equation}
	where $\mu_F$ is factorization scale and $\mu_R$ is renormalization scale. 
	The natural choice of $\mu_R$ in QED is the electron mass. 
	For $e^+ e^-$ annihilation into a $Z$ boson, $\mu_F$ can be chosen equal to the $Z$-boson mass. We use the standard modified minimal subtraction scheme $(\overline{\mathrm{MS}})$ for treatment of factorization. Other schemes like FKS~\cite{Engel:2019nfw} and DIS~\cite{Frixione:2012wtz} can be applied in the same way.
	
	The method of structure functions in QED was developed on the base 
	of the QCD parton distribution function approach. The
	Dokshitzer--Gribov--Lipatov--Altarelli--Parisi evolution equations were reduced to QED by E.A.~Kuraev and V.S.~Fadin~\cite{Kuraev:1985hb}. 
	There are numerous applications and further developments of the method
	within the leading logarithmic approximation, see e.g.
	Refs.~\cite{Nicrosini:1986sm,Przybycien:1992qe,Skrzypek:1992vk,Cacciari:1992pz,Jadach:2000ir,Arbuzov:2005pt,WorkingGrouponRadiativeCorrections:2010bjp}. 
	Application of the method in the next-to-leading order (NLO) approximation was for the first time demonstrated in~\cite{Berends:1987ab} 
	for derivation of QED radiative corrections due to the initial state radiation (ISR) in electron-positron annihilation. 
	Then it was applied for calculations of $\mathcal{O}(\alpha^2L)$ corrections
	to a few other processes including muon decay~\cite{Arbuzov:2002cn}, deep inelastic scattering~\cite{Blumlein:2002fy}, and Bhabha scattering~\cite{Arbuzov:2006mu}. Recently, the calculations of next-to-leading ISR
	corrections to $e^+e^-$ annihilation were extended to higher orders 
	up to $\mathcal{O}(\alpha^6 L^5)$~\cite{Ablinger:2020qvo}. 
	The details on derivation of the electron NLO
	parton distribution functions can be 
	found in~\cite{Bertone:2019hks,Arbuzov:2022fmv}.
	
	Because of the importance of higher order ISR corrections to $e^+e^-$ annihilation,
	we decided to perform an independent calculation of them. In particular, in~\cite{Arbuzov:2023qgc}
	we have already noticed a discrepancy in the $\mathcal{O}(\alpha^3 L^3)$ singlet contribution to 
	the electron PDF with respect to~\cite{Skrzypek:1992vk}. Below we will show the corresponding effect 
	in the ISR corrections. We also perform here a detailed comparison with the results presented in~\cite{Ablinger:2020qvo}. 
	%%\textbf{We suggest a factorization scheme to make our results agree with direct calculations and to avoid using the variable of integration in the factorization scale.}
	
	%The article is organized as follows. The next Section 
	
	\section{Calculations}
	
	\subsection{Master formula}
	
\begin{center}
		\begin{figure}[ht]
			\includegraphics[width=0.8\linewidth]{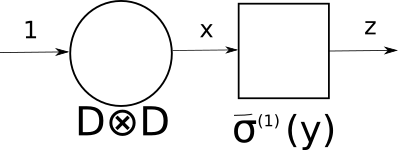}
			\caption{The scheme of energy fractions in the process.} \label{scheme}
		\end{figure}
	\end{center}

	The cross-section of electron-positron annihilation into a virtual photon or $Z$ boson $e^+e^- \to \gamma^{*}(Z^{*})$ can be represented in the form of convolution of two electron PDFs and partonic cross sections~\cite{Berends:1987ab}
	\begin{eqnarray} \label{master}
		&&	\sigma^{\mathrm{NLO}}_{\bar{e} e}(s') = \sum \limits_{i,j= e\!, \bar{e}\!,  \gamma}  \int \limits^{1}_{\bar{z_1}} \int \limits^{1}_{\bar{z_2}} d z_1 
		d z_2 D^{\mathrm{str}}_{i e} \left(z_1,\frac{\mu_R^2}{\mu^2_F}\right) 
		\nonumber \\
		&&	\times D^{\mathrm{str}}_{j \bar{e}} \! \left( \! z_2,\frac{\mu_R^2}{\mu^2_F} \! \right) \!
		\left( \sigma^{(0)}_{ij} (s z_1 z_2) + \bar{\sigma}^{(1)}_{ij} (s z_1 z_2) 
		+ \mathcal{O}(\alpha^2 L^0)  \right) \nonumber \\
		&&	\times \delta(s' \! - \! sz)
		+ \mathcal{O}\left(\frac{\mu_R^2}{\mu^2_F}\right),
	\end{eqnarray}
	where $\bar{e}\equiv e^+$ is positron and $e\equiv e^-$ is electron, $\sigma^{(0,1)}_{ij}$ are the Born $(0)$ and one-loop $(1)$ cross-sections of annihilation to $\gamma^*(Z^*)$ at the parton level, $s$ is the initial centre-of-mass energy squared, $s'$ is the invariant mass of the produced virtual photon (or $Z$-boson), $s'=sz$. 
	For $D \otimes D\otimes\sigma^{(0)}$ (see Fig.~\ref{scheme})
	$z = z_1z_2\equiv x$ because of the absence of the radiation in the Born-level partonic cross-section.
	In the case of one-loop contribution we have to introduce variable describing 
	possible energy losses due to radiation in the one-loop partonic cross-section. Let us assume that $y=z/x$ is the ratio of the squared invariant mass of the produced virtual photon to the squared invariant mass of colliding partons $i$ and $j$.
	So, the condition $s'=sz$ takes the form
	$s'=sz_1 z_2y= sxy =sz$.
	
	The process is schematically shown in Fig.~\ref{scheme}.
	
	The master formula for the cross-section in terms of convolutions from the evolution equation reads
	\begin{eqnarray} \label{eq:cross}
		&&\frac{d \sigma_{\bar{e} e}}{d s'} = \sigma^{(0)} \Bigl[ D_{e e} \otimes D_{\bar{e}  \bar{e} } \otimes {\sigma}_{e \bar{e } }  + D_{e e} \otimes D_{\gamma  \bar{e}} \otimes \sigma_{e \gamma} \nonumber \\
		&&+ D_{e e} \otimes D_{e \bar{e}} \otimes \sigma_{e e}  +  D_{\gamma e} \otimes D_{\bar{e} \bar{e}} \otimes\sigma_{\gamma \bar{e} }   \nonumber \\
		&&+ D_{\gamma e} \otimes D_{\gamma \bar{e}} \otimes \sigma_{\gamma \gamma }+ D_{\gamma e} \otimes D_{e \bar{e}} \otimes \sigma_{ \gamma e}   \nonumber \\
		&&+ D_{\bar{e} e} \otimes D_{\bar{e} \bar{e}} \otimes \sigma_{\bar{e} \bar{e}} 
		+ D_{\bar{e} e} \otimes D_{\gamma \bar{e}} \otimes \sigma_{\bar{e} \gamma}  \nonumber \\
		&&+D_{\bar{e} e}  \otimes D_{e \bar{e}} \otimes \sigma_{\bar{e} e} \Bigr],
	\end{eqnarray}
	where $D_{ia}$ are the parton distribution functions of parton $i$ in the initial particle $a$ and ${ \sigma}_{ij}$ are the partonic cross-sections, which in QCD are called  Wilson coefficients~\cite{Berends:1987ab}. The symbol $\otimes$ means convolution operation, 
	see e.g.~\cite{Arbuzov:2022fmv}.
	
	In these formulae all possible contributions to the NLO order are taken into account. In Table \ref{t1}, these contributions and their leading powers of $\alpha$ and the large logarithm are shown. In the Table, the symbol of convolution ($\otimes$) is omitted for convenience.
	
	In works~\cite{Berends:1987ab,Ablinger:2020qvo} only the following four contributions $D_{e e } \otimes D_{\bar{e} \bar{e}} \otimes \sigma_{e \bar{e}} $, $D_{e e} \otimes D_{\gamma  \bar{e}} \otimes \sigma_{e \gamma}$ , $ D_{\gamma e} \otimes D_{\bar{e} \bar{e}} \otimes\sigma_{\gamma \bar{e} } $ and $ D_{\gamma e} \otimes D_{\gamma \bar{e}} \otimes \sigma_{\gamma \gamma } $  were taken into account,
	i.e., the possibility to find positron in electron (an vice versa) were omitted. In paper~\cite{Berends:1987ab}, this limitation was well justified since the authors were interested only in $\mathcal{O}(\alpha^1)$ and $\mathcal{O}(\alpha^2)$ corrections to which the 
	electron-into-positron transitions do not contribute. Meanwhile, for higher-order corrections calculated in~\cite{Ablinger:2020qvo}, the transitions become relevant even in the leading logarithmic approximation.
	
	%\begin{center}
	\begin{table}[ht] 
		\caption{Orders of different contributions.} \label{t1}
		\begin{tabular}{|l|c|c|c|} 
			\hline
			\diagbox[width=3em]
			{i}{j} & $\bar{e}$ & $\gamma$ & $e$ \\
			\hline 
			$e$ & $D_{e e } D_{\bar{e} \bar{e}} \sigma_{e \bar{e}} $& $D_{ee} D_{\gamma\bar{e}} \sigma_{e \gamma}$  & $D_{e e} D_{e \bar{e}} \sigma_{e e}$ \\
			& LO (1) & NLO ($\alpha^2 L$)& NNLO ($\alpha^4 L^2$) \\
			\hline 
			$\gamma$ & $ D_{\gamma e} D_{\bar{e} \bar{e}} \sigma_{ \gamma \bar{e}} $ & $ D_{\gamma e} D_{\gamma \bar{e}} \sigma_{\gamma \gamma } $  & $ D_{\gamma e} D_{e \bar{e}} \sigma_{\gamma e} $\\
			& NLO ($\alpha^2 L$) & NNLO ($\alpha^4 L^2$) & NLO ($\alpha^4 L^3$) \\
			\hline 
			$\bar{e}$ & $D_{\bar{e} e} D_{\bar{e} \bar{e}} \sigma_{\bar{e} \bar{e}}$ & $ D_{\bar{e} e} D_{\gamma \bar{e}} \sigma_{\bar{e} \gamma}$ & $ D_{\bar{e} e} D_{e \bar{e}} \sigma_{\bar{e} e} $\\
			& NNLO ($\alpha^4 L^2$) & NLO ($\alpha^4 L^3$) & LO ($\alpha^4 L^4$)\\
			\hline
		\end{tabular}
	\end{table}
	%\end{center}
	
	\subsection{Evolution equations}
	
	Let us consider QED evolution equations for PDFs in the spacelike region. 
	The equations are induced by the renormalization group and have the following form, e.g., see~\cite{Arbuzov:2019hcg}:
	\begin{eqnarray} \label{evol_eq}
		&&	D_{ba}(x,\frac{\mu_R^2}{\mu^2_F}) = \delta(1-x)\delta_{ba} + \sum\limits_{i=e,\bar{e},\gamma}\int\limits_{\mu_0^2}^{\mu^2} 
		\frac{dt \alpha(t)}{2 \pi t} \nonumber \\
		&&\times \int\limits_{x}^{1} \frac{dy}{y} D_{ia}(y,t,\frac{\mu_R^2}{\mu^2_F}) P_{bi} \left( \frac{x}{y} \right), 
	\end{eqnarray}
	where index $a$ corresponds to the initial particle, e.g., an electron; and 
	indices $b$ and $i$ mark QED partons which can be photons $(\gamma)$ or massless 
	electrons $(e)$ and positrons $(\bar{e})$. Note that transition into
	all three types of partons have to be taken into account.
	
	Every splitting function, PDF or radiator can be divided into $\Theta$ and $\Delta$ parts as
	$$F(z) = \lim_{\Delta\to 0}\left(F_\Theta(z)\Theta(1-\Delta-z)+\delta(1-z)F_\Delta\right).$$
	Both appear in the process of $\Delta$-regularization of the functions divergent at $z \to 1$. The $\Theta$ part depends on the energy fraction $z$ and corresponds to hard photon or pair emission. The $\Delta$ part provides the contribution of virtual and soft radiation with the emitted energy fraction not more than $\Delta$. Details can be found in \cite{Arbuzov:2022fmv}.

	The splitting functions $P_{ji} (x)$ can be expanded in the fine structure constant $\alpha$
	\begin{eqnarray} \label{pee}
		P_{ji} (x) = P^{(0)}_{ji} (x) + \frac{\alpha(t)}{2 \pi} P^{(1)}_{ji} (x) + \mathcal{O}(\alpha^2).
	\end{eqnarray}
	
	We took the expressions for the NLO QCD splitting functions $P_{ji}^{(1)}$ from 
	Refs.~\cite{Ellis:1996nn,Furmanski:1980cm} for the spacelike case and from 
	Refs.~\cite{Furmanski:1980cm,Mele:1990cw} for the timelike ones. We reduced the
	functions to the (abelian) QED case by taking the appropriate values of constants: 
	$C_F=1$, $C_G = 0$, $T_R\cdot N_F = 1$ for \cite{Furmanski:1980cm} and $C_F=1$, $N_C = 0$, 
	and $T_f=1$ for \cite{Ellis:1996nn}. Note that the expressions for $P_{ij}^{(1)}$ must 
	be consistent with the running of $\alpha$ to avoid double counting, see Sect.~\ref{Sect_alpha} below. 
	Details of analytical iterative solution of evolution equations can be found in~\cite{Arbuzov:2022fmv}.

	%The evolution equations can be solved analytically using the method of iterations, see details in~\cite{Arbuzov:2022fmv}.
	
	The initial conditions for the QED DGLAP evolution equations can be found in, e.g., Refs.~\cite{Frixione:2019lga,Blumlein:2011mi,Arbuzov:2002cn}. 
	For the time-like case, one can use the abelian part of the initial conditions 
	for QCD fragmentation functions~\cite{Mele:1990cw}. 
	Details on derivation of the  $d_{ee}^{(1)}$ function were given in ~\cite{Andonov:2009nn}.
	
	\subsection{Factorization at NLO}
	
	The cross-section in the NLO approximation takes into account
	the QED radiative corrections enhanced by the large logarithms and reads
	\begin{eqnarray} \label{cij}
		d \sigma^{\mathrm{NLO}}_{ab\to cd} &=& d\sigma^{(0)}_{ab\to cd} \biggl\{ 1 + \sum^{\infty}_{k=1} 
		\left(\frac{\alpha}{2\pi}\right)^k \sum^k_{l=k-1} c_{kl} L^{l}  
		\nonumber \\ 
		&+& \mathcal{O}(\alpha^kL^{k-2}) \biggr\},
	\end{eqnarray}
	where $c_{kl}$ are the coefficients to be computed.
	The terms of the type $\alpha^k L^{k}$ provide the leading order (LO) logarithmic approximation, 
	and the ones of the type $\alpha^k L^{k - 1}$ yield the NLO contribution.
	
	So the expression for the one-loop correction to electron positron annihilation 
	(with reduced energy due to the initial state radiation) reads
	\begin{eqnarray} \label{delta_1_ee}
		&& \delta_{\bar{e} e}^{(1)} (sx)\equiv \frac{\sigma_{\bar{e} e}^{(1)} (sx)}{\sigma_{\bar{e} e}^{(0)} (sx)} =  \frac{\alpha}{\pi}  \biggl\{ \left[\frac{1+y^2}{1-y}\right]_{+} \left( \ln  \frac{s x}{m_e^2} - 1\right) \nonumber \\
		&& + \delta (1-y) \biggl( 2 \zeta_2 - \frac{1}{2} \biggr) \biggr\},\qquad y= \frac{z}{x},
	\end{eqnarray}
	and, analogously, 
	\begin{equation}
		\delta_{e \gamma}^{(0)} (sx)\equiv \frac{\sigma_{e \gamma}^{(0)} (sx)}{\sigma_{\bar{e} e}^{(0)} (sx)}.
	\end{equation}
	
	Formula~(\ref{delta_1_ee}) for $x=1$ represents one-loop ISR correction to the process
	of electron-positron annihilation for the center-of-mass energy $\sqrt{s}$. By looking at this 
	expression, we can see that the (\'a la Brodsky-Lepage-Mackenzie) factorization scale choice $\mu_F=\sqrt{s}$ 
	is well motivated, since it absorbs the bulk of the one-loop correction. So in our calculations, we adapt the latter factorization scale. In work~\cite{Berends:1987ab} and later in~\cite{Ablinger:2020qvo} the factorization
	scale $\mu_F=\sqrt{zs}$ was chosen, which is the invariant mass of the final state. The latter choice looks
	not optimal, especially for small $z$.

	This choice of $ \bar{\delta}_{\bar{e} e}^{(1)} (x)$ satisfies the matching equality
	\begin{equation} \label{matching}
		\delta^{(1)}_{\bar{e} e}(sx) = \bar{\delta}^{(1)}_{\bar{e} e}(sx) 
		+ 2 \frac{\alpha}{2\pi}\left[ P^{(0)}_{ee}(y) L + d_{ee}^{(1)}(y)\right] 
		+ \mathcal{O}\left(\frac{m_e^2}{s}\right),
	\end{equation}
	where on the left-hand side we have the known one-loop ISR correction~\cite{Berends:1987ab}.
	
	After the subtraction of mass singularities within the standard
	modified minimal subtraction scheme $(\overline{\mathrm{MS}})$, we get 
	\begin{eqnarray}
		&& \bar{\delta}_{\bar{e} e}^{(1)} (sx) = \frac{\alpha}{\pi} \biggl\{  \left[\frac{1+y^2}{1-y}\right]_{+} \left(  \ln z - \ln y\right) 
		\nonumber \\ && \quad
		+ 2 (1+y^2) \biggl[\frac{\ln (1-y) }{1-y}\biggr]_+  
		\nonumber \\ && \quad
		+ \delta (1-y) \biggl( 2 \zeta_2 - \frac{1}{2} \biggr)\biggr\} .
	\end{eqnarray}
	
	Note that "bar" over $\delta$ here means that the latter is calculated for massless partons.
	Note also that variable $z$ above is the energy fraction of the produced virtual photon or $Z$ boson, 
	and it is not a variable of integration. 
	
	In the works~\cite{Berends:1987ab,Ablinger:2020qvo} the factorization scale is 
	implicitly chosen as $\mu_F^2=sz$~\footnote{Such a choice is not justified by the known
		result for one-loop ISR corrections.}. So, the large logarithm in the electron PDFs 
	is $\ln(s/m_e^2) + \ln z$. But in the expression for the one-loop partonic cross section used in Refs.~\cite{Berends:1987ab,Ablinger:2020qvo}, variable $y=z/x$ was occasionally replaced by $x$. 
	So they had
	\begin{eqnarray}
		&& \left[\bar{\delta}_{\bar{e} e}^{(1)} (x)\right]^{*} = \frac{\alpha}{\pi} \biggl\{   \left[\frac{1+x^2}{1-x}\right]_{+} \ln x  + 2 (1+x^2) \biggl[\frac{\ln (1-x) }{1-x} \biggr]_+ \nonumber \\
		&& + \delta (1-x) \biggl( 2\zeta_2 - \frac{1}{2} \biggr) \biggr\}.
	\end{eqnarray}
	The result calculated with this deformation of the factorization scale choice occasionally coincides with the known result of 
	direct two-loop calculation in the leading and sub-leading logarithmic contributions~\cite{Berends:1987ab,Blumlein:2011mi}, 
	but in higher orders the two schemes give significantly different results. In particular, our  $\mathcal{O}(\alpha^3L^2)$ result for function $c_{32}(z)$ (Eq.~(\ref{eq:c32theta}) in the Appendix) considerably differs from the one given in~\cite{Ablinger:2020qvo}.

	The coefficient $c_{21}$, see Eq.~(\ref{cij}), calculated via solving the evolution equation
	in terms of convolutions contains the one-loop correction:
	\begin{eqnarray}
		&& c_{21} (z) = \sigma_{e \bar{e}}^{(0)} \biggl[ 2 \delta_{e \gamma}^{(0)} \otimes P_{\gamma e}^{(0)}       
		+ \frac{2}{3}\bar{\delta}_{\bar{e} e}^{(1)}                      	  + 2 \bar{\delta}_{\bar{e} e}^{(1)} \otimes  P_{ee}^{(0)}            
		\nonumber       \\ 
		&&	  + 2  P_{e \gamma}^{(0)}  \otimes d_{\gamma e}^{(1)}   + 2    P_{ee}^{(1)} - \frac{20}{9} P_{ee}^{(0)} + 4   P_{ee}^{(0)} \otimes d_{ee}^{(1)} \biggr].  
	\end{eqnarray}
	It agrees with the corresponding result in~\cite{Berends:1987ab}.
	And its $\Theta$ part as function of $z$ reads
	\begin{eqnarray}
		&&  c^\Theta_{21} (z) = \ln z   \Bigl(  - \frac{37}{6} - \frac{2}{3 z} + \frac{29}{6 (1-z)} + 2  \frac{\ln (1-z)}{1-z} + \ln (1-z) \nonumber \\ 
		&& 
		- \frac{11}{3} z + z \ln (1-z) + \frac{4}{3} z^2 \Bigr)    + \ln^2 z   \Bigl( \frac{1}{4} - \frac{2}{1-z} + \frac{1}{4} z \Bigr)
		+ \frac{2}{9}  \nonumber \\ 
		&&- \frac{8}{9 z}+ \frac{4}{3 z} \ln (1-z) - \frac{73}{9 (1-z)} + 2 \mathrm{Li}_2 (1-z)  - \frac{20}{3}  \frac{\ln (1-z)}{1-z} \nonumber \\
		\nonumber &&+ \frac{13}{3} \ln (1-z) +  \frac{4 \zeta_2}{1-z} - 2 \zeta_2 + \frac{71}{9} z + 2 z \mathrm{Li}_2 (1-z) \\
		&&+ \frac{7}{3} z \ln (1-z)  - 2 z \zeta_2 + \frac{8}{9} z^2 - \frac{4}{3} z^2 \ln (1-z),
	\end{eqnarray}
	and the $\Delta$ part is
	\begin{eqnarray}
		c^\Delta_{21} &=& -\frac{203}{12} +22 \zeta_2+12 \zeta_3-\frac{40 }{3} \ln^2 \Delta \nonumber \\ 
		& +&\biggl(16 \zeta_2-\frac{292}{9}\biggr) \ln \Delta .
	\end{eqnarray}

	%The complete result  to the order $\alpha^2 L^2$ in the terms of convolutions \begin{eqnarray}
		%&&\sigma^{NLO} = c_{10} + c_{11} + c_{21} +c_{22} =  \sigma_{e \bar{e}}^{(0)} 
		%      + \frac{\alpha}{2 \pi}  \biggl(  \sigma_{e \bar{e}}^{(1)}  
		%       +2  \sigma_{e \bar{e}}^{(0)}     d_{ee}^{(1)} \biggr) 
		%       + 2 \frac{\alpha}{2 \pi} L \sigma_{e \bar{e}}^{(0)}   P_{ee}^{(0)}  
		%         \nonumber \\
		%&& + \left( \frac{\alpha}{2 \pi} \right)^2 L   \biggl( 2 \sigma_{e \gamma}^{(0)}  P_{\gamma e}^{(0)}  + \frac{2}{3} %\sigma_{e \bar{e}}^{(1)}  
		%  + 2 \sigma_{e \bar{e}}^{(1)}  \otimes P_{ee}^{(0)} + \sigma_{e \bar{e}}^{(0)} \left(  2  d_{\gamma e}^{(1)} \otimes P_{e %\gamma}^{(0)}  + 2 P_{ee}^{(1)} - \frac{20}{9} P_{ee}^{(0)} \right. \nonumber \\
		%&&\left.  + 4 P_{ee}^{(0)} \otimes  d_{ee}^{(1)} \right)  \biggr) + \left( \frac{\alpha}{2 \pi} \right)^2 L^2 \sigma_{e %\bar{e}}^{(0)}    \Bigl( P_{\gamma e}^{(0)} \otimes P_{e \gamma}^{(0)}  + \frac{2}{3} P_{ee}^{(0)}  + 2 P_{ee}^{(0)} %\otimes P_{ee}^{(0)}  \Bigr).  
		%\end{eqnarray}

		\subsection{Running coupling} \label{Sect_alpha}
		
		We use the expression for the running coupling in the 
		$\overline{\mathrm{MS}}$ scheme 
		\begin{equation}
			\alpha(\mu^2) = \frac{{\alpha(\mu_0)}}{1 + \overline{\Pi}(\mu,\mu_0,\alpha(0))}\, ,
		\end{equation}
		that can be found, e.g., in \cite{Gorishnii:1991hw, Baikov:2012rr}
		with 
		\begin{eqnarray} \label{polar}
			\overline{\Pi}\left(\mu,\mu_0,\alpha(0)\right) &=&\frac{\alpha(0)}{\pi} \left( \frac{5}{9}-\frac{L}{3} \right) 
			+ \left(\frac{\alpha(0)}{\pi}\right)^2 \left( \frac{55}{48}-\zeta_3-\frac{L}{4} \right) \nonumber \\
			&+&\left(\frac{\alpha(0)}{\pi}\right)^3 \left( \frac{-L^2}{24} \right) + \ldots
		\end{eqnarray}
		where $L =\ln(\mu_R^2/\mu_F^2)$ is again the large logarithm.
		After expansion, we get
		\begin{eqnarray}
			&&	\alpha(\mu_R^2) =  \alpha (0) \biggl\{ 1 +\frac{\alpha (0)}{2 \pi}  
			\left(  - \frac{10}{9} + \frac{2}{3} L \right) 
			+ \left( \frac{\alpha (0)}{2 \pi} \right)^2    \nonumber \\ 
			&& \times	\biggl(  - \frac{1085}{324}+ 4 \zeta_3 - \frac{13}{27} L
			+ \frac{4}{9} L^2 \biggr)  + \mathcal{O} \left( \alpha^3(0)\right) \biggr\},
		\end{eqnarray}
		where $\zeta_n\equiv\zeta(n)$ is the Riemann zeta function.
		Here we put $\mu_R=m_e$ and assume $\alpha(\mu_R^2)\approx\alpha(0)\equiv\alpha$.
		
		Note that in the traditional way of $\overline{\mathrm{MS}}$ scheme application 
		in QCD calculations, the expansion for the running coupling constant takes into account 
		only the terms proportional to large logs (via $\beta_0$, $\beta_1$ and so on). 
		The effects due to constant (non-logarithmic) terms, like $-10/9$ of 
		the $\mathcal{O}(\alpha)$ order in the above formula, are kept in higher-order splitting functions, 
		e.g., in $P_{ij}^{(1)}$.
		Here we apply a QED-like scheme in which the non-logarithmic terms are preserved in
		the running $\alpha$ and thus we modify the NLO splitting functions in the following way:
		$\bigl[P_{ij}^{(1)}(x)\bigr]_{QED}=\bigl[P_{ij}^{(1)}(x)\bigr]_{QCD}+\frac{10}{9}P_{ij}^{(0)}(x)$, 
		see details in~\cite{Arbuzov:2022fmv}. One can verify that this scheme choice doesn't affect 
		the final results.

		\section{Results in terms of convolutions}
		
		Parton distribution functions of the types $D_{e \bar{e}}$ and $D_{\bar{e} e}$ start to give their contribution to cross-section from the order $\alpha^2 \ L^2$. 
		
		The complete results for $c_{33}$,  $c_{44}$, $c_{32}$, and $c_{43}$ in terms of convolutions read
		\begin{eqnarray}
			&&  c_{33} (z) =     \frac{2}{3} P_{e \gamma}^{(0)} \otimes P_{\gamma e}^{(0)}  + \frac{1}{3} P_{e \gamma}^{(0)} \otimes P_{\gamma e}^{(0)} \otimes P_{\gamma \gamma}^{(0)}  + \frac{8}{27}   P_{ee}^{(0)} \nonumber \\
			&& + \frac{5}{3} P_{ee}^{(0)} \otimes P_{e \gamma}^{(0)} \otimes P_{\gamma e}^{(0)}  + \frac{4}{3}  P_{ee}^{(0) \otimes 2} + \frac{4}{3}      P_{ee}^{(0) \otimes 3} ,
		\end{eqnarray}
		
		\begin{eqnarray}
			&&c_{44} (z) =   \sigma_{e \bar{e}}^{(0)} \biggl(  \frac{11}{27} P_{e \gamma}^{(0)} \otimes P_{\gamma e}^{(0)}   + \frac{1}{12} P_{e \gamma}^{(0)} \otimes P_{\gamma e}^{(0)} \otimes  P_{\gamma \bar{e}}^{(0)}   \nonumber \\
			&&    \otimes P_{\bar{e} \gamma}^{(0)}   + \frac{1}{3} P_{e \gamma}^{(0)} \otimes P_{\gamma e}^{(0)} \otimes P_{\gamma \gamma}^{(0)}  + \frac{1}{12} P_{e \gamma}^{(0)} \otimes P_{\gamma e}^{(0)} \otimes  P_{\gamma \gamma}^{(0) \otimes 2} \nonumber \\
			&&+ \frac{1}{4} 
			P_{e \gamma}^{(0) \otimes 2 } \otimes P_{\gamma \bar{e}}^{(0) \otimes 2}  + \frac{1}{3} P_{e \gamma}^{(0)\otimes 2} \otimes P_{\gamma e}^{(0)\otimes 2} + \frac{4}{27} P_{ee}^{(0)} \nonumber \\
			&&  + 
			\frac{5}{3} P_{ee}^{(0)} \otimes P_{e \gamma}^{(0)} \otimes P_{\gamma e}^{(0)}  + \frac{1}{2} P_{ee}^{(0)} \otimes P_{e \gamma}^{(0)} \otimes P_{\gamma e}^{(0)} \otimes P_{\gamma \gamma}^{(0)}   \nonumber \\
			&& + \frac{17}{12} P_{ee}^{(0)\otimes 2} \otimes P_{e \gamma}^{(0)}  \otimes P_{\gamma e}^{(0)}  + \frac{22}{27} 
			P_{ee}^{(0) \otimes 2}  \nonumber \\
			&&+ \frac{4}{3} P_{ee}^{(0)\otimes 3}+ \frac{2}{3} P_{ee}^{(0)\otimes 4} \biggr),
		\end{eqnarray}
		\begin{eqnarray}
			&&  c_{32} (z) =     \sigma_{e \bar{e}}^{(0)} \biggl\{ \delta_{e \gamma}^{(0)}  \otimes\biggl( P_{\gamma e}^{(0)} \otimes P_{\gamma \gamma}^{(0)}  + 3   P_{ee}^{(0)} \otimes P_{\gamma e}^{(0)}  \nonumber \\
			&& + 2 P_{\gamma e}^{(0)}    \biggr)  + \bar{\delta}_{\bar{e} e}^{(1)}  \otimes \biggl( P_{e \gamma}^{(0)} \otimes P_{\gamma e}^{(0)}  + 2  P_{ee}^{(0)} + 2 P_{ee}^{(0)} \otimes P_{ee}^{(0)} \biggr)  \nonumber \\
			&& + \frac{4}{9} \bar{\delta}_{\bar{e} e}^{(1)} +   P_{\gamma e}^{(0)} \otimes P_{e \gamma}^{(1)}  +  P_{e \gamma}^{(0)} \otimes P_{\gamma e}^{(1)}  + \frac{2}{3}  P_{e \gamma}^{(0)} \otimes d_{\gamma e}^{(1)}   \nonumber \\
			&&  - \frac{20}{9}  P_{e \gamma}^{(0)} \otimes P_{\gamma e}^{(0)} +  P_{e \gamma}^{(0)} \otimes d_{\gamma e}^{(1)} \otimes P_{\gamma \gamma}^{(0)}   + \frac{4}{3}  P_{ee}^{(1)}    \nonumber \\
			&& + 2 d_{ee}^{(1)} \otimes P_{e \gamma}^{(0)} \otimes P_{\gamma e}^{(0)} - \frac{13}{27}  P_{ee}^{(0)} + 3  P_{ee}^{(0)} \otimes  P_{e \gamma}^{(0)} \otimes  d_{\gamma e}^{(1)}   \nonumber \\
			&&  + 4 
			P_{ee}^{(0)} \otimes P_{ee}^{(1)}  + \frac{4}{3}  P_{ee}^{(0)} \otimes d_{ee}^{(1)}  - \frac{40}{9}  P_{ee}^{(0)} \otimes P_{ee}^{(0)} \nonumber \\
			&&   + 4 
			P_{ee}^{(0)} \otimes P_{ee}^{(0)} \otimes d_{ee}^{(1)} \biggr\},
		\end{eqnarray}
		
		\begin{eqnarray}
			&&c_{43} (z) =      \sigma_{e \bar{e}}^{(0)} \biggl\{ \delta_{e \gamma}^{(0)} \otimes \biggl(\frac{44}{27}  P_{\gamma e}^{(0)}  + \frac{1}{3}  P_{\gamma e}^{(0)} \otimes P_{\gamma \bar{e}}\otimes P_{\bar{e} \gamma} \nonumber \\
			&& + \frac{4}{3}  P_{\gamma e}^{(0)} \otimes P_{\gamma \gamma}^{(0)}  + \frac{1}{3}  P_{\gamma e}^{(0)} \otimes P_{\gamma \gamma}^{(0)} \otimes P_{\gamma \gamma}^{(0)} +   4  P_{ee}^{(0)} \otimes P_{\gamma e}^{(0)}  \nonumber \\
			&& + \frac{4}{3}  P_{e \gamma}^{(0)} \otimes P_{\gamma e}^{(0)} \otimes P_{\gamma e}^{(0)}  + \frac{4}{3}  P_{ee}^{(0)} \otimes P_{\gamma e}^{(0)} \otimes P_{\gamma \gamma}^{(0)} \nonumber \\
			&&  + \frac{7}{3}  P_{ee}^{(0)} \otimes P_{ee}^{(0)} \otimes 
			P_{\gamma e}^{(0)} \biggr) + \frac{8}{27} \bar{\delta}_{e\bar{e}}^{(1)}  + \bar{\delta}_{e\bar{e}}^{(1)} \biggl(\frac{4}{3} P_{e \gamma}^{(0)} \otimes P_{\gamma e}^{(0)}\nonumber \\
			&&   + \frac{1}{3}  P_{e \gamma}^{(0)} \otimes P_{\gamma e}^{(0)} \otimes P_{\gamma \gamma}^{(0)} + \frac{5}{3}  P_{ee}^{(0)} \otimes P_{e \gamma}^{(0)} \otimes P_{\gamma e}^{(0)}\nonumber \\
			&&          + \frac{44}{27}  P_{ee}^{(0)}   + \frac{8}{3}  P_{ee}^{(0)}  \otimes P_{ee}^{(0)}   + \frac{5}{6}  P_{\gamma e}^{(0)} \otimes P_{e \bar{e}}^{(1)} \otimes P_{\bar{e} \gamma}  \nonumber \\
			&&          + \frac{4}{3}  P_{ee}^{(0)\otimes 3}\biggr)  + \frac{10}{9} \sigma_{e \bar{e}}^{(0)} \otimes P_{\gamma e}^{(0)}  
			P_{e \gamma}^{(1)}  + \frac{1}{3}  P_{\gamma e}^{(0)} \otimes P_{\gamma \gamma}^{(0)}\otimes P_{e \gamma}^{(1)} \nonumber 
		\end{eqnarray}
		
		\begin{eqnarray}
			&&   + \frac{5}{6}  P_{e \gamma}^{(0)} \otimes P_{ \bar{e} \bar{e}}^{(1)} \otimes P_{\gamma \bar{e}}^{(0)} + \frac{8}{9} 
			P_{e \gamma}^{(0)} \otimes P_{\gamma e}^{(1)}  + \frac{8}{27}  P_{e \gamma}^{(0)} \otimes d_{\gamma e}^{(1)} \nonumber \\
			&& + \frac{1}{3}  P_{e \gamma}^{(0)} \otimes P_{\gamma e}^{(1)} \otimes P_{\gamma \gamma}^{(0)}   + \frac{5}{6}  P_{e \gamma}^{(0)} \otimes d_{\gamma e}^{(1)} \otimes P_{\gamma \bar{e}}^{(0)}  \otimes P_{\bar{e} \gamma} ^{(0)} \nonumber \\
			&&          + \frac{2}{3}  P_{e \gamma}^{(0)} \otimes d_{\gamma e}^{(1)} \otimes P_{\gamma \gamma}^{(0)} + \frac{1}{3} 
			P_{e \gamma}^{(0)} \otimes d_{\gamma e}^{(1)} \otimes P_{\gamma \gamma}^{(0)} \otimes P_{\gamma \gamma}^{(0)} \nonumber \\
			&&   - \frac{11}{9}  P_{e \gamma}^{(0)} \otimes P_{\gamma e}^{(0)}  + \frac{1}{3}  P_{e \gamma}^{(0)}  
			P_{\gamma e}^{(0)} \otimes P_{\gamma \gamma}^{(1)}   + \frac{8}{9}  P_{ee}^{(1)}\nonumber \\
			&&  - \frac{10}{9}  P_{e \gamma}^{(0)} \otimes P_{\gamma e}^{(0)} \otimes P_{\gamma \gamma}^{(0)}  + \frac{1}{2}  P_{e \gamma}^{(0)} \otimes P_{\gamma e}^{(0)} \otimes d_{\gamma e}^{(1)}  
			P_{\bar{e} \gamma} \nonumber \\
			&&  + \frac{4}{3}  P_{e \gamma}^{(0)} \otimes P_{e \gamma}^{(0)\otimes 2} \otimes d_{\gamma e}^{(1)}   + \frac{5}{3}   
			P_{ee}^{(1)} \otimes P_{e \gamma}^{(0)} \otimes P_{\gamma e}^{(0)}  \nonumber \\
			&& + \frac{4}{3}  d_{ee}^{(1)} \otimes P_{e \gamma}^{(0)} \otimes P_{\gamma e}^{(0)} + \frac{10}{81}  P_{ee}^{(0)}    + \frac{5}{3}  P_{ee}^{(0)} \otimes P_{\gamma e}^{(0)} \otimes P_{e \gamma}^{(1)}   \nonumber \\
			&& + \frac{2}{3}  d_{ee}^{(1)} \otimes P_{e \gamma}^{(0)} \otimes 
			P_{\gamma e}^{(0)} \otimes P_{\gamma \gamma}^{(0)}  + \frac{5}{3} 
			P_{ee}^{(0)} \otimes P_{e \gamma}^{(0)} \otimes P_{\gamma e}^{(1)}  \nonumber \\
			&&  + \frac{4}{3}  d_{\gamma e}^{(1)} \otimes P_{ee}^{(0)}  \otimes  P_{e \gamma}^{(0)} \otimes  P_{\gamma \gamma}^{(0)}  + 2  P_{ee}^{(0)} \otimes P_{e \gamma}^{(0)} \otimes d_{\gamma e}^{(1)} \nonumber \\
			&&  - \frac{50}{9}  P_{ee}^{(0)} \otimes P_{e \gamma}^{(0)} \otimes P_{\gamma e}^{(0)} + 4  P_{ee}^{(0)} \otimes P_{ee}^{(1)} + \frac{16}{27}  P_{ee}^{(0)} \otimes d_{ee}^{(1)} \nonumber \\
			&&   + \frac{10}{3}  P_{ee}^{(0)} \otimes d_{ee}^{(1)} \otimes P_{e \gamma}^{(0)} \otimes P_{\gamma e}^{(0)}  - \frac{22}{9} 
			P_{ee}^{(0) \otimes 2} + 4  P_{ee}^{(0) \otimes 2} \nonumber \\
			&&    \otimes P_{ee}^{(1)}  + \frac{7}{3}  P_{ee}^{(0) \otimes 2} \otimes P_{e \gamma}^{(0)} \otimes d_{\gamma e}^{(1)} + \frac{8}{3} 
			P_{ee}^{(0) \otimes 2} \otimes d_{ee}^{(1)}  \nonumber \\
			&&     + \frac{8}{3} \sigma_{e\bar{e}}^{(0)}  \otimes d_{ee}^{(1)}  \otimes P_{ee}^{(0) \otimes 3} \biggr\},
		\end{eqnarray}

		where $P_{ij}^{(0)\otimes n}$ is the successive convolution of $n$ $P_{ij}^{(0)}$ functions.
		There are no positron-induced contributions in $c_{32}$ because they appear only starting from the $\mathcal{O}(\alpha^4)$ order. The formula for $c{32}$ in terms of convolutions coincides with the one given in~\cite{Ablinger:2020qvo}.
		But we do not agree in the final result for this contribution (as a function of $z$) given in Appendix below, because of the difference in treatment of NLO factorization.
		
		In the $\mathcal{O}(\alpha^4L^4)$ contribution, 
		we have the difference with respect to the result from the work~\cite{Ablinger:2020qvo}
		\begin{eqnarray} \label{deltac44}
			&&\Delta c_{44} =    \frac{1}{3} \sigma_{e \bar{e}}^{(0)} P_{e \gamma}^{(0)} \otimes  P_{\gamma e}^{(0)} \otimes  P_{\gamma \bar{e}}^{(0)} \otimes  P_{\bar{e} \gamma}^{(0)} 
		\end{eqnarray}
		because of taking into account electron-into-positron transitions. 
		We have two sources of these transitions: including such transitions in evolution equations and including the term proportional to $D_{e \bar{e}} \otimes D_{\bar{e} e}$ in the master formula~(\ref{eq:cross}). If we exclude both parts, our result for $c_{44}$ completely coincides with the result from the work~\cite{Ablinger:2020qvo}. From the evolution equation, when we include the transitions into positrons in the equations for $D_{ee}$ and $D_{\gamma e}$, it comes $\frac{1}{12} \sigma_{e \bar{e}}^{(0)} P_{e \gamma}^{(0)} \otimes  P_{\gamma e}^{(0)} \otimes  P_{\gamma \bar{e}}^{(0)} \otimes  P_{\bar{e} \gamma}^{(0)}$, and the last term in Eq.~(\ref{eq:cross}) yields $\frac{1}{4} \sigma_{e \bar{e}}^{(0)} P_{e \gamma}^{(0)} \otimes  P_{\gamma e}^{(0)} \otimes  P_{\gamma \bar{e}}^{(0)} \otimes  P_{\bar{e} \gamma}^{(0)}$. 
		
\begin{figure}[ht] 
			\includegraphics[width=0.95\linewidth]{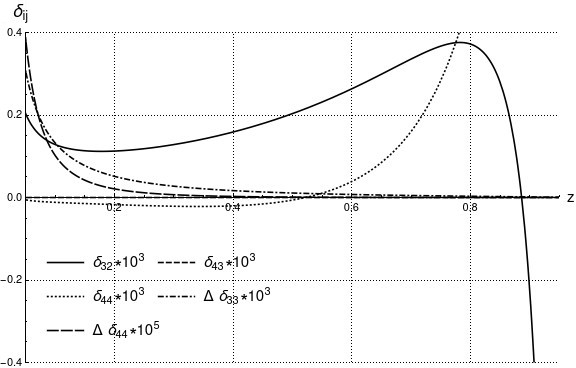} \label{C32334344}
			\caption{Higher-order contribution to the ISR radiator function.}
		\end{figure}

		Numerical illustrations of our results in the orders $\mathcal{O}(\alpha^3 L^2)$, $\mathcal{O}(\alpha^4 L^4)$ 
		and $\mathcal{O}(\alpha^4 L^3)$ are shown in Fig.~\ref{C32334344}. We plot the values
		\begin{equation}
			\delta_{ij} = \left(\frac{\alpha}{2\pi}\right)^iL^j\frac{c_{ij}}{\sigma_{e\bar{e}}^{(0)}}
		\end{equation}
		as functions of $z$ (we put $L=24$, i.e., $\mu_F\approx M_Z$). Note that these quantities are contributions to the
		ISR radiator functions which have to be later integrated with the Born level cross section
		over $z$ in an interval defined by the experiment. 
		We also show the difference $\Delta\delta_{33}$ in the $\mathcal{O}(\alpha^3 L^3)$ order, which comes from the correction in the singlet part of $D_{ee}$ with respect to the result given in~\cite{Skrzypek:1992vk}, and the difference $\Delta\delta_{44}$ between our fourth-order leading logarithmic contribution $\delta_{44}$ and the one from~\cite{Ablinger:2020qvo}, which is due to the electron-into-positron transitions. 
		One can see that all shown contributions are relevant for future experiments with the precision tag
		of the order $10^{-5}$. 
		The radiator function contributions typically diverge for $z\to 1$, but taking into account 
		$\Delta$ parts cancels out this divergence in the total correction.
		The differences $\Delta\delta_{33}$ and $\Delta\delta_{44}$ are enhanced at small $z$ since
		both are related to singlet transitions.

		\section{Conclusions}
		
		In this way, we revisited the application of the QED structure function method for calculations
		of higher-order NLO ISR radiative corrections to $e^+e^-$ annihilation. 
		A bug in the singlet part of the $\mathcal{O}(\alpha^3L^3)$ of
		the electron PDF is corrected. 
		
		Several other issues that arose in earlier calculations of these corrections are clarified
		and improved. The relevance of positron in electron (and vice versa) 
		PDF is demonstrated explicitly. Taking into account splitting of electron into
		positron (and vice versa) appears to be also significant in solutions of the QED evolution equations. 
		Moreover, treatment of the factorization scale in NLO was refined. 
		The issues listed above lead to a considerable difference of our 
		results (both the leading and next-to-leading ones) from the ones given in~\cite{Ablinger:2020qvo}.
		
		%With the help of a complete $\mathcal{O}(\alpha^2)$ result, e.g., for the same process 
		%of electron-positron annihilation with initial
		%state radiation corrections~\cite{Berends:1987ab, Blumlein:2011mi},
		%one can fix the next-to-leading contributions to the splitting functions and
		%the subtracted one-loop cross-sections $\bar{\sigma}^{(1)}_{ij\to kl}$.
		
		The obtained results will be implemented into the {\tt ZFITTER} computer code~\cite{Arbuzov:2005ma}.
		We would like to underline that the applied method leads to results being integrated over angular variables of the 
		ISR radiation, and hence they can not take into account all possible experimental cuts.
		Nevertheless, first of all, one can use our results as benchmarks to verify the precision of Monte Carlo codes.
		Moreover, after implementation of the completely differential two-loop radiative corrections in a Monte Carlo
		code, one can add certain higher-order corrections in the collinear approximation
		without spoiling the theoretical precision too much.
		Extension of the presented here results to higher orders, like $c_{55}$ and $c_{54}$ is straightforward.
		The corresponding results will be presented elsewhere.

		\begin{acknowledgments}
			We are grateful to prof. V.S.~Fadin for useful discussions. A.A. thanks the Russian Science Foundation for the support (project No. 22-12-00021).
			
		\end{acknowledgments}
		
		\appendix*
		
		\section{Results as functions of $z$} \label{appendix:fz}
		
		Here, we show explicit results for the higher-order coefficients $c_{kl}$ from Eq.~(\ref{cij}) 
		as functions of the energy fraction $z$.
		
		\begin{eqnarray}
			&& c^\Theta_{33} = -\frac{461}{18}-\frac{7 z}{18}+\frac{710}{27
				(1-z)}+\frac{16}{27 z}+\zeta_2 \biggl(16 z-\frac{32}{1-z} \nonumber \\
			&& +16\biggr)+\biggl(\frac{44 z}{3}+\frac{44}{3}\biggr) \mathrm{Li}_2(1-z)+\frac{8 z^2}{27 (1-z)}-\frac{16
				z^2}{27}\nonumber \\
			&& +\biggl(\frac{40 z^2}{9}+\frac{128 z}{9}-\frac{88}{3 (1-z)}+\frac{242}{9}\biggr) \ln (z) \nonumber \\
			&&+ \biggl[-\frac{40 z^2}{9}-\frac{50
				z}{3}+\frac{176}{3 (1-z)}+\frac{40}{9 z}+\biggl(\frac{92 z}{3} -\frac{32}{1-z}+\frac{82}{3}\biggr)\nonumber \\
			&& \times \ln (z)-42\biggr] \ln (1-z) +\biggl(-16 z+\frac{32}{1-z}-16\biggr) \ln ^2(1-z)\nonumber \\
			&&+\biggl(-\frac{19 z}{3}+\frac{16}{3 (1-z)}-\frac{19}{3}\biggr) \ln
			^2(z)
		\end{eqnarray}
		
		\begin{eqnarray}
			&&c^\Theta_{44} (z) = -\frac{18779}{432}+ \ln^3(1-z) \left(-\frac{64 z}{3}+\frac{128}{3
				(1-z)}-\frac{64}{3}\right)  \nonumber \\
			&& \quad +\ln(z) \biggl[\ln(1-z)^2 \left(\frac{178
				z}{3}-\frac{64}{1-z}+\frac{178}{3}\right)+\ln(1-z) \nonumber \\
			&& \quad \times \left(\frac{136
				z^2}{9}+\frac{194 z}{3}-\frac{128}{1-z}+\frac{352}{3}\right)+\zeta_2
			\biggl(-\frac{32 z}{3} \nonumber 
		\end{eqnarray}
	
         \begin{eqnarray}
			&& \quad +\frac{64}{3 (1-z)}  -\frac{32}{3}\biggr)+\mathrm{Li}_2 (1-z)
			\left(-38 z+\frac{64}{3 (1-z)}-38\right) \nonumber \\
			&& \quad +\frac{64 z^2}{27}  -\frac{241
				z}{72}-\frac{1708}{27 (1-z)}-\frac{32}{27
				z}+\frac{1145}{24}\biggr]+\biggl(-\frac{68 z^2}{9}  \nonumber \\
			&& \quad -\frac{113
				z}{3}+\frac{128}{1-z}+\frac{68}{9 z}-\frac{271}{3}\biggr) \ln^2(1-z) +
			\biggl[ \biggl(-\frac{146 z}{3}  \nonumber \\
			&& \quad +\frac{128}{3
				(1-z)}-\frac{146}{3}\biggr)\ln(1-z)-\frac{34 z^2}{9}-\frac{275 z}{24}+\frac{64}{3
				(1-z)}\nonumber \\
			&& \quad -\frac{187}{8}\biggr] \ln^2(z) +\ln(1-z) \biggl[\zeta_2 \left(64
			z-\frac{128}{1-z}+64\right)+ \biggl(\frac{164
				z}{3}  \nonumber \\
			&& \quad +\frac{164}{3}\biggr) \mathrm{Li}_2 (1-z)-\frac{32 z^2}{9}-\frac{659 z}{54}+\frac{3416}{27
				(1-z)}+\frac{32}{9 z}-\frac{11173}{108}\biggr] \nonumber \\
			&& \quad +\ln(z)^3 \left(\frac{61
				z}{36}-\frac{16}{9 (1-z)}+\frac{61}{36}\right)+\zeta_2 \biggl(\frac{68
				z^2}{9}+\frac{113 z}{3}  \nonumber \\
			&& \quad -\frac{128}{1-z}-\frac{68}{9
				z}+\frac{271}{3}\biggr)+\zeta_3 \left(6 z+\frac{128}{3
				(1-z)}+6\right)+\biggl(\frac{68 z^2}{9}   \nonumber \\
			&& \quad +27 z+\frac{68}{9
				z}+27\biggr) \mathrm{Li}_2 (1-z) +\mathrm{Li}_3 (z) \biggl(-\frac{146 z}{3}+\frac{128}{3
				(1-z)}   \nonumber \\
			&& \quad -\frac{146}{3}\biggr)+\mathrm{Li}_3 (1-z) \left(-\frac{164
				z}{3}-\frac{164}{3}\right)+\frac{328 z^2}{81}+\frac{1115 z}{432}   \nonumber \\
			&& \quad +\frac{368}{9
				(1-z)}-\frac{328}{81 z},
		\end{eqnarray}

		\begin{eqnarray}
			&&\Delta c^\Theta_{44} (z) =  \frac{1}{3} \biggl[ \ln z   \left(  - 21 - \frac{16}{9 z} - 21 z - \frac{16}{9} z^2 \right)   +  \ln^2 \! z  \ (  - 2 + 2 z )   \nonumber \\
			&&    +  \ln^3 \!  z  \!  \left( \!  - \frac{2}{3} - \frac{2}{3} z \!  \right) \!   - 26 - \frac{176}{27 z} + 26 z + \frac{176}{27} z^2  \!  \biggr].
		\end{eqnarray}
		
		\begin{eqnarray}
			&& c^\Theta_{32} (z) = \frac{1453}{9}-\frac{935 z^2}{54
				(1-z)}-\frac{413 z^2}{27}-\frac{910 z}{9}-\frac{4967}{54 (1-z)} \nonumber \\
			&& \quad +\frac{584}{27
				z}+\ln(z) \biggl(\ln(1-z)^2 \biggl(10 z+\frac{32}{1-z}+10\biggr)+
			\biggl(\frac{104 z^2}{3}\nonumber \\
			&& \quad -\frac{154 z}{3}+\frac{224}{3 (1-z)}-\frac{8}{3
				z}-\frac{328}{3}\biggr)\ln(1-z)+\zeta_2 \biggl(48
			z \nonumber \\
			&& \quad -\frac{128}{1-z}+48\biggr)+\mathrm{Li}_2(1-z) \biggl(-28
			z+\frac{32}{1-z}-28\biggr) \nonumber 	\\
			&& \quad -\frac{100 z^2}{9} -\frac{1597 z}{18}+\frac{214}{3
				(1-z)}-\frac{20}{9 z}+\frac{67}{9}\biggr)+ \biggl(-\frac{52
				z^2}{3}  \nonumber \\
			&& \quad +19 z-\frac{64}{1-z}+\frac{52}{3 z}+45\biggr)\ln^2(1-z)+
			\biggl(\ln(1-z) \biggl(-4 z   \nonumber \\
			&& \quad -\frac{16}{1-z} -4\biggr)-\frac{20 z^2}{3}+\frac{16
				z}{3}-\frac{92}{3 (1-z)}+\frac{185}{6}\biggr) \ln^2(z) + \nonumber 
			\end{eqnarray}

           \begin{eqnarray}
			&& \quad \biggl(\frac{88 z^2}{9
				(1-z)}  +\frac{136 z^2}{9}+\frac{1678 z}{9}-\frac{1768}{9 (1-z)}-\frac{100}{9
				z}+\frac{142}{9}  \nonumber \\
			&& \quad + \zeta_2
			\biggl(-32 z +\frac{64}{1-z}-32\biggr)+\mathrm{Li}_2(1-z) (52 z+52)\biggr)\nonumber \\
			&& \quad \times  \ln (1-z) +\ln^3(z) \biggl(-\frac{23 z}{3}+\frac{32}{3
				(1-z)}-\frac{23}{3}\biggr)\nonumber \\
			&& \quad +\zeta_2 \biggl(\frac{20 z^2}{3 (1-z)} +\frac{64
				z^2}{3} -\frac{40 z}{3}+\frac{292}{3 (1-z)}-\frac{64}{3
				z}-\frac{220}{3}\biggr) \nonumber \\
			&& \quad +\zeta_3 \biggl(\frac{24 z^2}{1-z} + 40
			z-\frac{40}{1-z}+40\biggr)+\mathrm{Li}_2(1-z) \biggl(\frac{64 z^2}{3}\nonumber \\
			&& \quad -40
			z+\frac{32}{3 z}-46\biggr) +\mathrm{Li}_3(z) \biggl(-40
			z+\frac{64}{1-z}-40\biggr)\nonumber \\
			&& \quad +\mathrm{Li}_3(1-z) (-52 z-52),
			\label{eq:c32theta}
		\end{eqnarray}
		
		\begin{eqnarray}
			&&c^\Theta_{43}=  \frac{191465}{648}+
			\biggl(\frac{211 z}{36}-\frac{16}{3 (1-z)}+\frac{191}{36}\biggr) \ln ^4(z) \nonumber \\
			&& \quad + \biggl(\frac{232 z^2}{27}-\frac{316 z}{27} +\biggl(-\frac{100
				z}{3}+\frac{160}{3 (1 \!- \! z)}-\frac{140}{3}\biggr) \ln (1 \!- \! z)\nonumber \\
			&& \quad +\frac{376}{9 (1-z)}-\frac{1727}{54}\biggr) \ln ^3(z) +\biggl(16 z^2+\frac{727
				z}{6}+\biggl(-\frac{176 z}{3}\nonumber \\
			&& \quad  -32\biggr) \ln ^2(1-z)+\biggl(\frac{40 z}{3}-\frac{40}{3}\biggr) \ln ^2(1+z) +\zeta_2\biggl(-86
			z  \nonumber \\
			&& \quad  +\frac{448}{3 (1-z)}-66\biggr)+\biggl(-\frac{800 z^2}{9}+\frac{508 z}{9}-\frac{368}{3 (1-z)} +\frac{1444}{9}\nonumber \\
			&& \quad -\frac{80}{9 z}\biggr) \ln
			(1-z)+\biggl(-\frac{40 z^2}{9}+\frac{10 z}{3}+\frac{10}{3}-\frac{40}{9 z}\biggr) \ln (1+z)\nonumber \\
			&& \quad-\frac{2780}{27 (1-z)}+134.833\, +\frac{88}{27
				z}\biggr) \ln ^2(z)+\biggl(\biggl(12 z+\frac{64}{1-z}\nonumber \\
			&& \quad +12\biggr) \ln ^3(1-z)+\biggl(\frac{952 z^2}{9}-\frac{874 z}{9}+\frac{688}{3
				(1-z)}-\frac{2476}{9} \nonumber \\
			&& \quad -\frac{112}{9 z}\biggr) \ln ^2(1-z)+\biggl(-\frac{1720 z^2}{27}-\frac{18248 z}{27}+\zeta_2\biggl(208
			z \nonumber \\
			&& \quad -\frac{576}{1-z}+\frac{544}{3}\biggr) +\frac{12100}{27 (1-z)}  -\frac{4556}{27}-\frac{296}{27 z}\biggr) \ln (1-z) \nonumber \\
			&& \quad +\biggl(\frac{40 z}{9}-\frac{40}{9}\biggr) \ln ^2(1+z)-\frac{3650
				z^2}{81}+\zeta_3\biggl(-68 z+\frac{32}{1-z} \nonumber \\
			&& \quad -\frac{284}{3}\biggr)+\frac{14267
				z}{54}+\zeta_2\biggl(-\frac{464 z^2}{9} +142 z-\frac{1472}{3 (1-z)}+\frac{1318}{3} \nonumber \\
			&& \quad +\frac{64}{3 z}\biggr)+\biggl(-\frac{40 z^2}{3}+\frac{320
				z}{9}+\zeta_2\biggl(\frac{160}{3}-\frac{160 z}{3}\biggr)+\frac{320}{9} \nonumber \\
			&& \quad -\frac{40}{3 z}\biggr) \ln (1+z)+\frac{6947}{27
				(1-z)}+\frac{344}{27 z}-\frac{3541}{36}\biggr) \ln (z)+\biggl(\frac{80 z}{3} \nonumber
			\end{eqnarray}

         \begin{eqnarray}
			&& \quad -\frac{80}{3}\biggr) \ln (1+z) \mathrm{Li}_2(1+z) \ln
			(z)+\biggl(-\frac{88 z^2}{3}+\frac{62 z}{3} \nonumber \\
			&& \quad  -\frac{256}{3 (1-z)}+\frac{194}{3}+\frac{88}{3 z}\biggr) \ln ^3(1-z)+\biggl(\frac{80
				z^2}{27}-\frac{20 z}{9}-\frac{20}{9} \nonumber \\
			&& \quad +\frac{80}{27 z}\biggr) \ln ^3(1+z)-\frac{3259 z^2}{162 (1-z)}-\frac{245 z^2}{9}+\biggl(\frac{584
				z^2}{27}  \nonumber \\
			&& \quad +\frac{7675 z}{18}+\zeta_2\biggl(-64 z+\frac{128}{1-z}-64\biggr)-\frac{1376}{3 (1-z)}+\frac{341}{18}  \nonumber \\	
	      	&& \quad -\frac{224}{27 z}\biggr) \ln
	      	^2(1-z)+\biggl(-\frac{136 z}{3}+\frac{64}{1-z}-\frac{56}{3}\biggr) \mathrm{Li}_2(1-z){}^2  \nonumber	\\
			&& \quad +\biggl(\frac{10}{3}-\frac{10 z}{3}\biggr)
			\mathrm{Li}_2\biggl(z^2\biggr){}^2+\zeta_2^2 \biggl(-\frac{728 z}{3}+\frac{448}{3 (1-z)}-\frac{728}{3}\biggr) \nonumber \\
			&& \quad -\frac{61121 z}{648}+\zeta_4
			\biggl(\frac{1468 z}{3}-\frac{2080}{3 (1-z)}+506\biggr)+\zeta_3\biggl(-\frac{56 z^2}{3 (1-z)}  \nonumber \\
			&& \quad -\frac{80 z^2}{9}-\frac{56 z}{3}-\frac{392}{3
				(1-z)}+52+\frac{848}{9 z}\biggr)+\zeta_2\biggl(\frac{1292 z^2}{27 (1-z)}  \nonumber \\
			&& \quad -\frac{104 z^2}{9}-\frac{10799 z}{27}+\frac{12524}{27
				(1-z)}-\frac{1001}{27}+\frac{104}{9 z}\biggr)  \nonumber \\
			&& \quad +\biggl(\frac{572 z^2}{27 (1-z)}+\frac{3026 z^2}{81}-\frac{3805 z}{27}+\zeta_3\biggl(32
			z-\frac{64}{1-z}+32\biggr)  \nonumber \\
			&& \quad +\zeta_2\biggl(-\frac{64 z^2}{3 (1-z)}+\frac{752 z^2}{9}-\frac{340 z}{3}+\frac{1376}{3
				(1-z)}-\frac{1100}{3}  \nonumber \\
			&& \quad-\frac{752}{9 z}\biggr)-\frac{5680}{9 (1-z)}+755.741\, -\frac{290}{81 z}\biggr) \ln (1-z) \nonumber \\
			&& \quad +\zeta_2\biggl(-\frac{80
				z^2}{9}+\frac{20 z}{3}+\frac{20}{3}-\frac{80}{9 z}\biggr) \ln (1+z)+\biggl(-\frac{1568 z^2}{27} \nonumber \\
			&& \quad -263 z+(132 z+132) \ln
			^2(1-z)+\biggl(-\frac{20 z}{3}-20\biggr) \ln ^2(z) \nonumber \\
			&& \quad +\zeta_2\biggl(-80 z-\frac{320}{3}\biggr)+\biggl(\frac{1072 z^2}{9}-\frac{1172
				z}{9}-\frac{1400}{9} +\frac{608}{9 z}\biggr) \nonumber \\
			&& \quad \times \ln (1-z)+\biggl(-\frac{688 z^2}{9}+\frac{232 z}{9}+\biggl(-\frac{544
				z}{3}+\frac{128}{1-z} \nonumber \\
			&& \quad - 128\biggr) \ln (1-z)+\frac{256}{3 (1-z)}-\frac{152}{9}-\frac{64}{3 z}\biggr) \ln (z)-\frac{1321}{9} \nonumber \\
			&& \quad -\frac{88}{27
				z}\biggr) \mathrm{Li}_2(1-z)+\biggl(-\frac{32 z}{9}-\frac{32}{9}\biggr) \mathrm{Li}_2(z)+\biggl(-\frac{20 z^2}{3} \nonumber \\
			&& \quad +\frac{160 z}{9}+\biggl(\frac{10
				z}{3}-\frac{10}{3}\biggr) \ln ^2(z)+\zeta_2\biggl(\frac{20 z}{3}-\frac{20}{3}\biggr)+\biggl(\frac{40 z^2}{9}  \nonumber \\
			&& \quad +\frac{20
				z}{3}+\biggl(\frac{40}{3}-\frac{40 z}{3}\biggr) \ln (1-z)-\frac{20}{3}-\frac{40}{9 z}\biggr) \ln (z)+\biggl(\frac{40}{3}  \nonumber \\
			&& \quad -\frac{40
				z}{3}\biggr) \mathrm{Li}_2(1-z)+\frac{160}{9}-\frac{20}{3 z}\biggr) \mathrm{Li}_2\biggl(z^2\biggr)+\biggl(-\frac{1072 z^2}{9}  \nonumber \\
			&& \quad +\frac{1172
				z}{9}+(-264 z-264) \ln (1-z)+\biggl(\frac{80 z}{3}+\frac{128}{1-z}+\frac{80}{3}\biggr)
			\nonumber \\
			&& \quad \times \ln (z)+\frac{1400}{9}-\frac{608}{9 z}\biggr)
			\mathrm{Li}_3(1-z)+\biggl(-\frac{368 z^2}{9}-\frac{70 z}{3}  \nonumber 
								\end{eqnarray}
		
	\begin{eqnarray}	
			&& \quad +\biggl(-128 z+\frac{256}{1-z}-128\biggr) \ln (1-z)+\biggl(-\frac{268
				z}{3}-116\biggr)   \nonumber \\
			&& \quad \times \ln (z) +\frac{512}{3 (1-z)}-\frac{286}{3}-\frac{80}{9 z}\biggr) \mathrm{Li}_3(z)+\biggl(-\frac{20 z^2}{3}  \nonumber \\
			&& \quad -5
			z+\biggl(\frac{10}{3}-\frac{10 z}{3}\biggr) \ln (z)+\frac{25}{3}+\frac{20}{9 z}\biggr) \mathrm{Li}_3\biggl(z^2\biggr)
			 \nonumber \\	
			&& \quad +\biggl(-\frac{160
				z^2}{9}+\frac{40 z}{3} +\biggl(\frac{80}{3}-\frac{80 z}{3}\biggr) \ln (z)+\frac{40}{3}-\frac{160}{9 z}\biggr)  \nonumber \\
			&& \quad \times 
			\mathrm{Li}_3\biggl(\frac{1}{1+z}\biggr)+(264 z +264) \mathrm{Li}_4(1-z)+\biggl(\frac{776 z}{3} \nonumber \\
			&& \quad +\frac{776}{3}\biggr) \mathrm{Li}_4(z)+\biggl(\frac{124
				z}{3}  +\frac{124}{3}\biggr) \mathrm{S}_{2,2}(z)-\frac{35281}{162 (1-z)} \nonumber \\
			&& \quad +\frac{641}{27 z}.
		\end{eqnarray}
		
		\begin{eqnarray}
			&& c_{33}^{\Delta} = \biggl( \frac{718}{27} - 32 \zeta_2 \biggr) \ln \Delta +     \frac{88}{3} \ln^2 \Delta +   \frac{32}{3} \ln^3 \Delta   + \frac{143}{18} \nonumber \\
			&& \quad + \frac{64}{3} \zeta_3 - \frac{88}{3} \zeta_2, 
		\end{eqnarray}
		
		\begin{eqnarray}
			&& c_{44}^{\Delta}  =    \biggl( \frac{1112}{27} + \frac{256}{3} \zeta_3 - 128 \zeta_2 \biggr)  \ln \Delta +    \biggl( \frac{1708}{27} - 64 \zeta_2 \biggr) 
			\nonumber \\
			&& \quad \times \ln^2 \Delta +    \frac{128}{3}  \ln^3 \Delta + \frac{32}{3} \ln^4 \Delta     + \frac{715}{72} + 16 \zeta_4 + \frac{256}{3} \zeta_3 \nonumber \\
			&& \quad - \frac{1708}{27} \zeta_2, \\
			&& c_{32}^{\Delta}  =    \biggl(  - \frac{2951}{27} + 48 \zeta_3 + 104 \zeta_2 \biggr) \ln \Delta +   \biggl( 32 \zeta_2 - \frac{280}{3} \biggr) \ln^2 \Delta  \nonumber \\
			&&  \quad - \frac{64}{3} \ln^3 \Delta      - \frac{1387}{36} - 80 \zeta_4 + \frac{4}{3} \zeta_3 + \frac{928}{9} \zeta_2, \\
			&& c_{43}^{\Delta}  = 
			-\frac{64 \ln^4 \Delta}{3}+\ln^3 \Delta \left(\frac{128
				\zeta_2}{3}-\frac{1376}{9}\right)+\ln^2 \Delta \biggr(\frac{656
				\zeta_2}{3}  \nonumber \\
			&& \quad + 96 \zeta_3-\frac{8234}{27}\biggr)+\ln \Delta
			\left(\frac{13816 \zeta_2}{27}+\frac{64 \zeta_3}{3}-320
			\zeta_4-\frac{19270}{81}\right) \nonumber \\
			&& \quad+\zeta_2 \left(\frac{8521}{27}-\frac{32
				\zeta_3}{3}\right)-\frac{1894 \zeta_3}{9}-\frac{776
				\zeta_4}{3}-\frac{15265}{216} .
		\end{eqnarray}
		
		Here $H (a_1, ..., a_k;z)$ are harmonic polylogarithms~\cite{Maitre:2005uu, Ablinger:2018sat}. 
		
		\bibliography{C32-V2_arxiv}% Produces the bibliography via BibTeX.
		
	\end{document}